\documentclass[twocolumn,secnumarabic,amssymb, nobibnotes, aps, prd]{revtex4-1}

\usepackage{graphicx}
\usepackage{hyperref}
\hypersetup{colorlinks,citecolor=blue}
\usepackage{color}
\begin{document}

\title{Physical implications of pure Lovelock geometry on stellar structure}%

\author{Ksh. Newton Singh}%
\email[Email:]{ntnphy@gmail.com}
\affiliation{Department of Physics, National Defence Academy, Khadakwasla, Pune-411023, India. \\
Department of Mathematics, Jadavpur University, Kolkata-700032, India}

\author{Megandhren Govender}%
\email[Email: ]{megandhreng@dut.ac.za}
\affiliation{Department of Mathematics, Faculty of Applied Sciences, Durban University of Technology, Durban, South Africa.}

\author{ Sudan Hansraj}%
\email[Email: ]{hansrajs@ukzn.ac.za}
\affiliation{Astrophysics and Cosmology Research Unit, School of Mathematics, Statistics and Computer Science, University of KwaZulu-Natal, Private Bag X54001, Durban 4000, South Africa.}

\author{Farook Rahaman}%
\email[Email:]{rahaman@associates.iucaa.in}
\affiliation{Department of Mathematics, Jadavpur University, Kolkata-700032, India}

\date{August 10, 2010}%

\begin{abstract}

 We construct an exact anisotropic  star model with a linear barotropic equation of state and with  Finch-Skea potential within the framework of pure Lovelock gravity. A comparison with the corresponding Einstein model in a suitable limit  is easily deduced. Evidently higher curvature effects induced by the Lovelock contributions generate lower densities, pressures, surface tensions and anisotropy factors when compared to its Einstein counterpart. The maximum moment of inertia is attained for the Einstein case and hence it may be inferred that Lovelock effects soften the equation of state. The model satisfies various stability tests.    
\end{abstract}


\maketitle

\section{INTRODUCTION}

Massive stars and supernovae have long been proposed as providing suitable conditions for important physical processes such as nucleosynthesis. Much depends on the physical properties of such stars. In particular its density, compactification, mass-radius relationships, gravitational surface redshifts, moments of inertia and equations of state furnish information on how such stellar laboratories operate. Presently, calculations and deductions from observations have principally been undertaken in the framework of general relativity (GR). However, there is a growing body of evidence suggesting that GR needs modification. If this is the case then there will be profound effects on the way most  physical processes have been understood to date. In this work, we investigate for the first time  what the physical consequences would be for stellar distributions if pure Lovelock  geometry were invoked instead of GR. The advantage of this approach is that the models we develop reduce to  GR since GR is the four dimensional version of first order Lovelock gravity thus facilitating direct comparisons using identical parameter values.

Experimental data suggests that the standard theory of gravity GR  may require  modifications. For example that the universe is currently undergoing an epoch of accelerated expansion has been amply confirmed by  the Hubble space telescope, Supernovae Ia as well as the WMAP surveys. These results are unexpected according to the standard theory. In order to address this shortcoming, conjectures of exotic matter fields have been made. It is proposed that about $75 \%$  of the energy budget of the universe consists of mysterious dark energy and about $27 \%$ is dark matter,  however, no experimental support for their existence has been forthcoming.   For recent theoretical work on dark energy and dark matter see \cite{tenkanen,bar,calmet}.

A completely different approach is to modify the theory of gravity presently in use. Einstein's theory has enjoyed notable successes including satisfying solar system tests, the confirmed prediction of gravitational waves \cite{GW}  and most recently the shadow of a black hole as investigated by the Event Horizon Telescope \cite{EHT}. Therefore, any modifications to the theory should preserve these positive features. Moreover any theory of gravity should preserve diffeomorphism invariance or the satisfaction of the Bianchi identities. There are a variety of such modified theories on the market including $f(R, T)$ \cite{harko,hans-ban}, Rastall theory \cite{Rastall,Rastall1}, unimodular (trace--free gravity) \cite{unimod1,unimod2,unimod3,ellis1,ellis2,hans-ellis}, massive gravity\cite{derham} and so on.  

Does there exist a theory that is diffeomorphism invariant and generates up to second order field equations and does not violate the energy conditions? The answer is in the affirmative. It is the Lovelock theory which to zeroth order corresponds to a vacuum solution with a cosmological constant, to first order gives the standard Einstein equations and to second order generates the Einstein--Gauss--Bonnet (EGB)  theory of gravity. The EGB theory finds support from string theory as the very same Lagrangian appears in the low energy effective action of heterotic string theory \cite{gross}. The Lovelock polynomial is the tensor-only action that  generates  up to second order equations of motion. If the effects of a scalar field is incorporated with a tensor field then we must use the Horndeski \cite{horndeski}  modification. The full blown EGB equations are notoriously complicated and only a few exact solutions for compact objects have been reported in the literature as opposed to the some 120 exact solutions known for the standard Einstein theory. The exterior spacetime for EGB theory was worked out by Boulware and Deser \cite{boulware} and its extension to involve the electromagnetic field was accomplished by Wiltshire \cite{wiltshire}.
Note that  Lovelock  higher curvature effects are active only in spacetime dimensions higher than 4.  It is unresolved how to explain how the extra dimensions are topologically hidden. Note however, that higher dimensional spaces are demanded in string theory and its generalization $M-$ theory. Moreover gravitational fields with higher dimensions were studied from the time of Kaluza and Klein \cite{kaluza,klein}  when the coupling of electromagnetic field to ordinary matter proved challenging and the extra dimension was explained away as a topological compactification.  A recent speculative idea advanced by Glavan and Lin \cite{glavan} introduced the scaling of the coupling constant by a factor of $D - 4$, $D$ being the spacetime dimension, in order to generate nontrivial curvature effects in the four dimensional paradigm. This proposal suffers the as yet unsettled problem of a discrete variable such as $D$ behaving as a continuous variable in approaching 0.

A study of the full Lovelock equations is prohibitive. Nevertheless it is possible to investigate the effects of extra curvature of arbitrary order by isolating particular terms of the Lovelock polynomial. This is the essence of this article. This study will provide some insight on how higher order Lovelock terms influence the gravitational behaviour of hyperspheres of anisotropic astrophysical fluids. The question of how to analyse the physical properties of higher dimensional objects arises since there exists no experimental or observational data.   Then it follows that the only way to connect with observables is to invoke data pertaining to well studied four dimensional objects. This is the path taken in this article.

\section{FIELD EQUATIONS IN Pure LOVELOCK GRAVITY}

The $N^{th}$ order Lovelock polynomial action is defined by the Lagrangian \cite{lov71}
\begin{eqnarray}
\mathcal{L}=\sum_{N=0}^N \alpha_N \mathcal{R}^{(N)} 
\end{eqnarray}
where
\begin{equation}
\mathcal{R}^{(N)}={1 \over 2^N}~\delta^{\mu_1 \nu_1......\mu_N \nu_N}_{\alpha_1 \beta_1....\alpha_N \beta_N} ~\prod_{r=1}^N R^{\alpha_r \beta_r}_{\mu_r \nu_r}~.
\end{equation}

Here $R^{\alpha \beta}_{\mu \nu}$ is the generalized Riemann tensor in $N^{th}$ order Lovelock gravity \cite{dad10}. Also the $\delta^{\mu_1 \nu_1......\mu_N \nu_N}_{\alpha_1 \beta_1....\alpha_N \beta_N}= {1\over N!} \delta^{\mu_1}_{\alpha_1} \delta^{\nu_1}_{\beta_1}.....\delta^{\mu_N}_{\alpha_N} \delta^{\nu_N}_{\beta_N}$ is the generalized Kronecker delta.\\

On varying the action including the Lagrangian density of the matter with respect to the metric, we obtain the equations of motion given by
\begin{eqnarray}
T_{AB} &=& \sum_{N=0}^N \alpha_N \mathcal{G}^{(N)}_{AB} \nonumber\\
 &=& \sum_{N=0}^N \alpha_N \Big[N\Big(\mathcal{R}^{(N)}_{AB}-{1 \over 2} \mathcal{R}^{(N)} g_{AB}\Big)\Big]
\end{eqnarray}

where $\mathcal{G}^{(N)}_{AB}$ is the $N^{th}$ order Einstein tensor, $\mathcal{R}^{(N)}=g^{AB}\mathcal{R}^{(N)}_{AB}$ and $T_{AB}$ is the energy-momentum tensor. For $N=0$ the gravitational equation in Lovelock gravity corresponds to the  cosmological constant, $N=1$ to Einstein's equation and $N=2$ to Gauss-Bonnet gravity, etc. The $N^{th}$ term of the equation of motion can be written as
\begin{eqnarray}
\mathcal{G}^{(N)}_{AB}=N\Big(\mathcal{R}^{(N)}_{AB}-{1 \over 2} \mathcal{R}^{(N)} g_{AB}\Big)=T_{AB} ~.
\end{eqnarray}

Now let us assume the interior space-time to be spherically symmetric and $d-$dimensional, and is in the form 
\begin{eqnarray}
ds^2=-e^\nu dt^2+e^\lambda dr^2+r^2 d\Omega^2_{d-2} \label{int}
\end{eqnarray}
where $d\Omega^2_{d-2}$ is the metric on a unit $(d-2)-$sphere and is given by 

\begin{eqnarray}
d\Omega_1^2 &=& d\phi^2 \nonumber \\
d\Omega_{i+1}^2 &=& d\theta_i^2+ \sin ^2 \theta_i ~d\Omega_i^2~,~~~i \ge 1. \nonumber \\
\end{eqnarray}

Assuming the comoving fluid velocity vector $u^i=e^{-\nu/2}\delta_0^i$, the corresponding stress tensor of an anisotropic fluid is $T_i^j=\mbox{diag} (-\rho,~p_r,~p_\theta,~p_{\phi},~~p_{\phi 1}...)$. The conservation of stress tensor i.e. $T_{A;B}^B=0$ yields the generalized TOV-equation in $d-$dimensions i.e.
\begin{eqnarray}
-{1 \over 2} (p_r+\rho){d\nu \over dr}-{dp_r \over dr}+{d-2 \over r}(p_\theta-p_r)=0~. \label{tov}
\end{eqnarray}
This equation can also be written in terms of the balanced force equation given by 
\begin{eqnarray}
F_g+F_h+F_a=0 \label{forc}
\end{eqnarray}
where $F_g$ is the gravitational force, $F_h$ is the hydrostatic force and $F_a$, the anisotropic force and defined as
\begin{eqnarray}
F_g &=& -{1 \over 2} (p_r+\rho){d\nu \over dr} \label{fg}\\
F_h &=& -{dp_r \over dr}\\
F_a &=& {d-2 \over r}(p_\theta-p_r)~. \label{fa}
\end{eqnarray}

The  density, radial \& tangential pressure can be expressed as
\begin{eqnarray}
\rho(r) &=& {(d-2)
e^{-\lambda}(1-e^{-\lambda})^{N-1} \over 2r^{2N} } \times \nonumber\\
&& \Big[rN\lambda'+\{d-2N-1\}\{e^\lambda-1\} \label{den}\Big]\\
p_r(r) &=& {(d-2)e^{-\lambda}(1-e^{-\lambda})^{N-1} \over 2r^{2N} } \times \nonumber\\
&& \Big[rN\nu'-\{d-2N-1\}\{e^\lambda-1\}\Big]\label{pr}\\
p_\theta(r) &=& {e^{-\lambda N}(e^\lambda-1)^{N-2} \over 4r^{2N} } \Big[Nr\lambda' \big\{2(d-2N-1) \nonumber \\
&& \{1-e^\lambda \}-r\nu'(e^\lambda-2N+1) \big\}-(1-e^\lambda) \nonumber\\
&& \Big\{2(d-2N-1)(d-2N-2)\{1-e^\lambda\}+2N  \nonumber\\
&& (d-2N-1)r\nu'+Nr^2\nu'^2  +2Nr^2 \nu'' \Big\}\Big] \label{pt}
\end{eqnarray}
respectively. 
Here primes ($'$) represent differentiation with respect to $r$ and we have set the Lovelock coupling parameter $\alpha_N$ to unity. Note that we are assuming an anisotropic fluid distribution  $p_r \neq p_{\theta}$ for the purpose of devising a realistic model. Anisotropy in self-gravitating systems has been explored in the seminal work of Herrera and Santos \cite{anis}. Some sources of local anisotropy include viscosity, pion condensation, neutrino trapping and rotation amongst others. In systems where the central density is of the order of $10^{11}$ to $10^{12} g.cm^{-3}$, the long mean free path of trapped neutrinos within the core results in a small radiative Reynolds number, thus rendering the stellar fluid viscous. For a rotating star, the slow-rotation approximation shows small deviations from spherical symmetry. The condition for hydrostatic equilibrium in the first approximation for slowly rotating fluids is the same as that of an anisotropic fluid. 
As to the question of what mechanisms give rise to attractive or repulsive forces due to anisotropy, the answer lies within the physics of the model. In the case of rotating fluids in the slow-rotation approximation, it can be shown that the anisotropy is related to the angular velocity $\omega$ via $\Delta=p_\theta-p_r=  \rho_0 \omega^2 r/3$.  
It is clear from this relation that the tangential pressure dominates the radial pressure, thus giving rise to a repulsive force due to anisotropy. Also, when the anisotropy is related to an outgoing null fluid (dissipation in the streaming out approximation), or when anisotropy is related to viscosity, one can show that the radial pressure dominates the tangential pressure. In the case of phase transitions, superconductivity, superfluidity, magnetic field, etc., the nature of the force due to anisotropy depends on the specific set up of the problem.

\section{A bounded SOLUTION IN pure LOVELOCK GRAVITY}

To solve the above sets of coupled differential equations (\ref{den})-(\ref{pt}), we  assume a linear equation of state (EoS) of the form
\begin{equation}
p_r(r)=\alpha \rho(r)-\beta \label{eos}
\end{equation}
with $\alpha$ and $\beta$ are arbitrary constants with appropriate units. To ensure a subluminal EOS, the velocity of sound $v_r=dp_r/d\rho=\alpha$ must be less than or equal to unity. Hence $\alpha$ should be chosen in the range $0  \leq \alpha \leq  1$. This EOS may be recognised as the MIT bag model for $\alpha=1/3$ and $\beta=4\mathcal{B}_g/3$, where $\mathcal{B}_g$ is the Bag constant.

Using (\ref{den}) and (\ref{pr}), (\ref{eos}) reduces to 
\begin{eqnarray}
\nu' &=& {1 \over rN} \Big[\alpha rN \lambda'-{2\beta r^{2N} \over d-2}~e^\lambda(1-e^{-\lambda})^{1-N}\Big] \nonumber\\
&& +\frac{(\alpha +1) (e^\lambda-1) (d-2 N-1)}{rN}. \label{dnu}
\end{eqnarray} 

To integrate (\ref{dnu}), we prescribe a particular form of $g_{11}$ of the Finch-Skea \cite{fin89} type metric potential given by
\begin{equation}
e^\lambda = 1+{r^2 \over R^2} \label{elam}
\end{equation}
where $R$ is an arbitrary parameter measured in $km$. This choice is motivated by the pleasing physical properties of the Finch-Skea model and its consistency with the astrophysical theory of Walecka \cite{walecka}. Walecka devised a relativistic mean field theory using a numerical integration of the main field equation for neutron stars. When the exact model of Finch and Skea was compared to the Walecka calculations it was shown that only an average of about a 12\%  deviation in the central mass density and surface density was evident. 

\begin{figure}[htbp]
\includegraphics[scale=0.7]{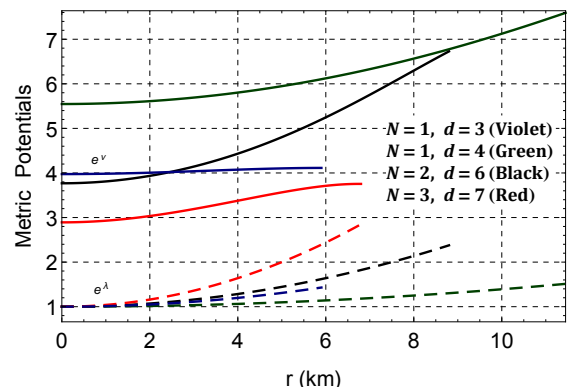}
\caption{\label{met}  Variation of metric potentials with radial coordinate $r$ for the parameters given in Table \ref{tab}.  }
\end{figure}

On using (\ref{elam}) in (\ref{dnu}) we get
\begin{eqnarray}
\nu(r) &=& \nu_0+\frac{(\alpha +1) r^2 (d-2 N-1)}{2 N R^2}+\alpha  \ln \left[R^2 \left\{1+{r^2 \over R^2} \right\}\right] \nonumber \\
&&  -\frac{\beta  r^{2 N+2} }{(d-2) N (N+1) R^2}~\left(\frac{r^2/R^2}{1+r^2/R^2}\right)^{-N-1} \label{nu}
\end{eqnarray}
where $\nu_0$ is the constant of integration. The variation of these metric functions are shown in Fig. \ref{met}.

\begin{figure}[htbp]
\includegraphics[scale=0.7]{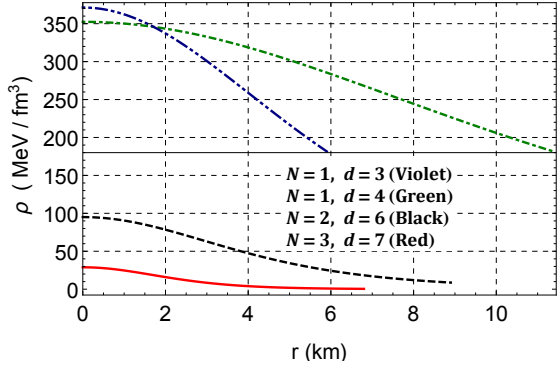}
\caption{\label{den}  Variation of density with radial coordinate $r$ for the parameters given in Table \ref{tab}.}
\end{figure}

The expressions of density and pressures can be written as
\begin{eqnarray}
\rho(r) &=& \frac{d-2}{2}  r^{-2 (N+1)} \left(\frac{r^2/R^2}{1+r^2/R^2}\right)^{N+1} \times \nonumber\\
&& \Big[r^2 (d-2 N-1)+(d-1) R^2\Big]\\
p_r(r) &=& \frac{\alpha(d-2)}{2}  r^{-2 (N+1)} \left(\frac{r^2/R^2}{1+r^2/R^2}\right)^{N+1} \times \nonumber\\
&& \Big[r^2 (d-2 N-1)+(d-1) R^2\Big]-\beta\\
p_\theta(r) &=& \frac{r^{4-2 N}(r^2/R^2)^{n-2}}{4 (d-2)^2 N R^{10} (1+r^2/R^2)^2} \left(1+\frac{r^2}{R^2}\right)^{-N}  \nonumber\\
&& \Big[2 \{2-d\} f_3(r) N R^4 \{d-2N-1\} (1+r^2/R^2) \nonumber\\
&& -2 \{2-d\} f_4(r) N R^2+2 \{2-d\} \Big(2 \{d-2\} \nonumber\\
&&  N R^4 \{d-2 N-1\} \{1+r^2/R^2\}-f_5(r)\{r^2 \nonumber\\
&& -2 (N-1) R^2\}\Big)N R^2+f_1(r)+f_2(r) r^2\Big].
\end{eqnarray}
The anisotropy is defined by $\Delta=p_\theta-p_r$. 

\begin{figure}[htbp]
\includegraphics[scale=0.7]{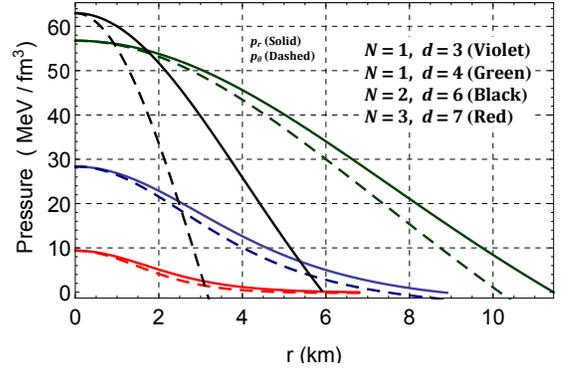}
\caption{\label{pres}  Variation of radial and transverse pressure with radial coordinate $r$ for the parameters given in Table \ref{tab}.  }
\end{figure}

\begin{figure}[htbp]
\includegraphics[scale=0.7]{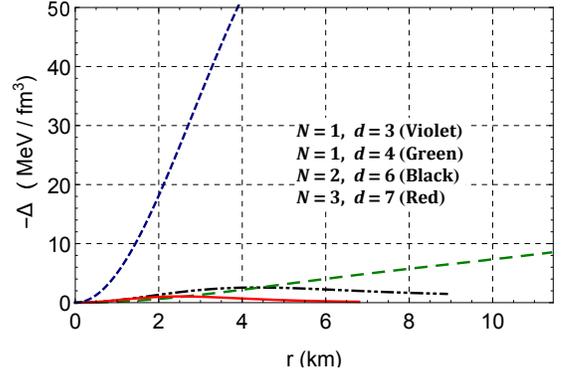}
\caption{\label{ani}  Variation of anisotropy with radial coordinate $r$ for the parameters given in Table \ref{tab}.}
\end{figure}

The variations of density, pressures and anisotropy are shown in Figs. \ref{den}, \ref{pres} and \ref{ani}. It can be seen that for $(N=1,~d=3)$ the $(\rho_c,~p_c,\Delta(R))$ are their highest values while in $(N=3,~d=7)$ they attain their lowest values. This signifies that the pressure, density and anisotropy decrease in higher order gravity and higher dimensions dimensions.

Here
\begin{eqnarray}
f_1(r) &=& -2 (d-2)^2 N R^6 (d-2 N-1) [d-2 (N+1)] \nonumber\\
&& (1+r^2/R^2)^2\\
f_2(r) &=& \bigg[(\alpha +1) (d-2) r^2 (d-2 N-1)+(d-2) R^2 \nonumber\\
&& (\alpha  d-\alpha +d-2 N-1)-2 \beta  r^{2 N+2} \nonumber\\
&& \left(\frac{r^2/R^2}{1+r^2/R^2}\right)^{-N} -2 \beta  R^2 r^{2 N} \nonumber \\
&& \left(\frac{r^2/R^2}{1+r^2/R^2}\right)^{-N}\bigg]^2\\
f_3(r) &=& -(\alpha +1) (d-2) r^2 (d-2 N-1)-(d-2) R^2 \nonumber\\
&& (\alpha  d-\alpha +d-2 N-1)+2 \beta  r^{2 N+2}  \nonumber\\
&& \left(\frac{r^2/R^2}{1+r^2/R^2}\right)^{-N}+2 \beta  R^2 r^{2 N} \nonumber \\
&& \left(\frac{r^2/R^2}{1+r^2/R^2}\right)^{-N}\\
f_4(r) &=& (\alpha +1) (d-2) r^4 (d-2 N-1)+2 (d-2) r^2 R^2 \nonumber\\
&& (\alpha  d-\alpha +d-3 \alpha  N-2 N-1)+(d-2) R^4 \nonumber\\
&& (\alpha  d-\alpha +d-2 N-1)-2 \beta  (2 N+1) r^{2 N+4} \nonumber\\
&& \left(\frac{r^2/R^2}{1+r^2/R^2}\right)^{-N}-4 \beta  (N+1) R^2 r^{2 N+2}  \nonumber\\
&& \left(\frac{r^2/R^2}{1+r^2/R^2}\right)^{-N}-2 \beta  R^4 r^{2 N} \left(\frac{r^2/R^2}{1+r^2/R^2}\right)^{-N}\\
f_5(r) &=& -(\alpha +1) (d-2) r^2 (d-2 N-1)-(d-2) R^2 \nonumber\\
&& (\alpha  d-\alpha +d-2 N-1)+2 \beta  r^{2 N+2}  \nonumber\\
&& \left(\frac{r^2/R^2}{1+r^2/R^2}\right)^{-N}+2 \beta  R^2 r^{2 N} \left(\frac{r^2/R^2}{1+r^2/R^2}\right)^{-N}
\end{eqnarray}

\begin{figure}[htbp]
\includegraphics[scale=0.7]{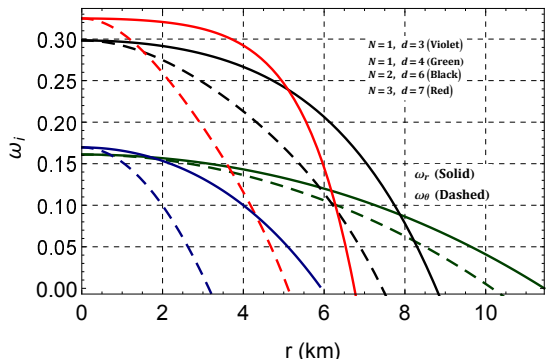}
\caption{\label{eos}  Variation of equation of state parameter ($\omega_r=p_r/\rho$ and $\omega_t=p_t/\rho$) with radial coordinate $r$ for the parameters given in Table \ref{tab}.  }
\end{figure}

Now we can define the equation of state parameters as
\begin{equation}
\omega_r = {p_r \over \rho}~~,~~\omega_\theta = {p_\theta \over \rho}.
\end{equation}
These parameters are less than unity implying that the solution can represent physically reasonable matter distributions (see Fig. \ref{eos}).

The density and pressure gradients can be expressed as
\begin{eqnarray}
{d\rho \over dr} &=& - \frac{(d-2) N r^{1-2 N} \left[r^2 (d-2 N-1)+(d+1) R^2\right]}{R^4\left(1+r^2/R^2\right)^2} \nonumber\\
&& \times \left(\frac{r^2/R^2}{1+r^2/R^2}\right)^N\\
{dp_r \over dr} &=& - \frac{(d-2) N r^{1-2 N} \left[r^2 (d-2 N-1)+(d+1) R^2\right]\alpha}{R^4\left(1+r^2/R^2\right)^2} \nonumber\\
&& \times \left(\frac{r^2/R^2}{1+r^2/R^2}\right)^N.
\end{eqnarray}
We omit writing the expression for $dp_\theta/dr$ due to its cumbersome and lengthy form.

\section{PROPERTIES OF THE SOLUTION}

\begin{figure}[htbp]
\includegraphics[scale=0.8]{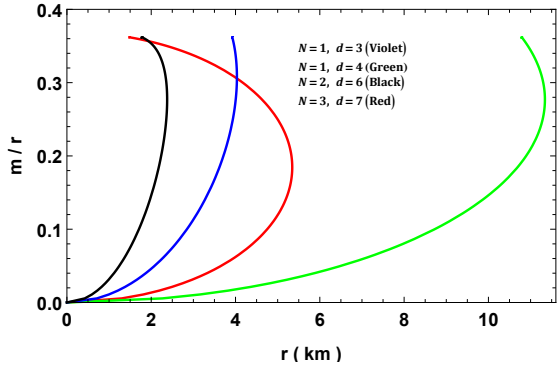}
\caption{\label{com}  Variation of compactness function with radial coordinate $r$ for the parameters given in Table \ref{tab}.   }
\end{figure}

The non-singular nature of the solution can be confirmed by checking the values of the  density and pressure at the center of the star. Furthermore,  the metric functions are regular at the centre. Both metric functions $e^{\nu}$ and $e^{\lambda}$ are monotonically increasing functions of $r$ as evidenced in Fig. \ref{met}. Fig. \ref{den} shows that the density decreases from the centre  towards the stellar surface. It is interesting to observe that the density decreases as one transcends from classical Einstein gravity through to pure Einstein-Gauss-Bonnet gravity to higher order pure Lovelock gravity. The behaviour in the radial pressure follows the trend of the energy density as one expects from the imposition of an equation of state as shown in Fig. \ref{pres}. The tangential pressure mimics the behaviour of the radial pressure at each interior point of the stellar configuration. We also observe that higher order effects result in smaller compact objects (i.e. smaller radii). The equation of state parameter is displayed in Fig. \ref{eos}. We observe  that the equation of state parameter increases with $N$ and $d$ thus allowing us to conclude that higher order effects coupled with higher dimensional effects lead to more compact objects. This is corroborated by Fig. \ref{com}, as will be examined later. The anisotropy increases towards the surface of the star Fig \ref{ani}. Higher order effects seem to quench the anisotropy at each interior point of the stellar configuration. 

The central values of density and pressure can we written as
\begin{eqnarray}
\rho_c &=& \frac{(d-2) (d-1)}{2 R^{2 N}} > 0~,~~\forall~~R>0\\
p_{rc} &=& p_{\theta c} = \frac{\alpha (d-2) (d-1)}{2 R^{2 N}}-\beta>0. \label{pc}
\end{eqnarray}

The  stiffest EoS known in   current physics is the Zeldovich's fluid where $p=\rho$ and hence the speed of sound  is equal to that of light. Therefore, the maximum possible ratio of pressure to density is unity for Zeldovich fluids and less than 1 for any other fluid. Even at the center  any stellar system composed of ordinary fluid must satisfy the Zeldovich's criterion given by
\begin{equation}
{p_{rc} \over \rho_c} = \alpha - \frac{2 R^{2 N}\beta}{(d-2) (d-1)} \le 1. \label{zel}
\end{equation} 

Using the above constraints (\ref{pc}) and (\ref{zel}) we can deduce the relationship 
\begin{eqnarray}
0<\left[\frac{\alpha (d-2) (d-1)}{2 R^{2 N}}-\beta\right] \le \frac{(d-2) (d-1)}{2 R^{2 N}}.
\end{eqnarray} 

\begin{figure}[htbp]
\includegraphics[scale=0.7]{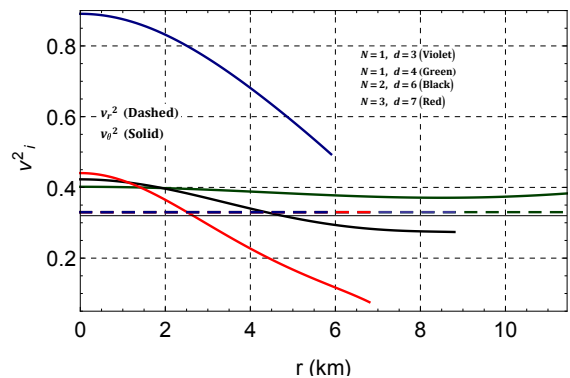}
\caption{\label{soun}  Variation of sound speed square with radial coordinate $r$ for the parameters given in Table \ref{tab}.   }
\end{figure}

The speed of sound in the interior can be determined as
\begin{equation}
v_r^2 = {dp_r \over d\rho}=\alpha~~,~~v_\theta^2={dp_\theta \over d\rho}.
\end{equation}
For physically plausible stellar configurations the causality condition demands that $0 \le v_r^2 \le 1$ and $0 \le v_\theta^2 \le 1$ (Fig. \ref{soun}). 

The non-singular nature of the central sound speed can also be seen from
\begin{eqnarray}
v_{rc} &=& \alpha\\
v_{\theta c} &=& \frac{R^{4 N}}{2 (d-2)^3 (d+1) N^2} \bigg[\frac{(d-2)^2}{R^{4 N}} \Big\{d^2 \big[1+\alpha ^2- \nonumber\\
&& 2 \alpha  (N^2-1)\big]+2 d \Big\{(\alpha +1)^2+\alpha  N^2+ \nonumber\\
&& (\alpha +1) N\Big\}+(\alpha +1) (\alpha +2 N+1)- \nonumber\\
&& \frac{4 \beta  (d-2)}{R^{2 N}} \Big\{\alpha -(\alpha +1) d+N+1\Big\}-4\beta^2 \bigg].
\end{eqnarray}
The compactness and mass confined in a radius $r$ of any stellar fluid configuration in $d-$dimensions is given by
\begin{eqnarray}
u(r) &=& {m(r) \over r^{d-3}} \label{comp}\\
m(r) &=& {2\pi^{(d-1)/2} \over \Gamma \left(\frac{d-1}{2}\right)}\int_0^r\rho(\xi) ~\xi^{d-2}d\xi . \label{mrs}
\end{eqnarray}

\begin{figure}[htbp]
\includegraphics[scale=0.7]{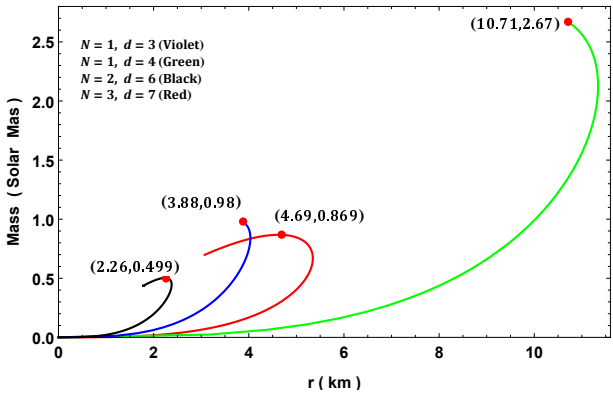}
\caption{\label{mas}  Variation of $M-R$ curve for the parameters given in Table \ref{tab}.  }
\end{figure}

Using the above definitions (\ref{comp}) and (\ref{mrs}) we determine the compactness parameter and mass function as
\begin{eqnarray}
&& u(r) = \frac{2 \pi ^{(d-1)/2}}{\Gamma \left(\frac{d-1}{2}\right)} \left[\frac{d-2}{16 \pi } r^{2-2 N} \left(\frac{r^2/R^2}{1+r^2/R^2}\right)^N\right] \label{comp1}\\
&& \hspace{-0.5cm} m(r) = \frac{2 \pi ^{(d-1)/2}}{\Gamma \left(\frac{d-1}{2}\right)} \left[\frac{d-2}{16 \pi } r^{d-2 N-1} \left\{\frac{r^2/R^2}{1+r^2/R^2}\right\}^N\right]\label{mrs1}
\end{eqnarray}
It may be observed from Fig. \ref{mas} that $m(0) = 0$ and the mass is an increasing function of the radial coordinate. In addition, lower mass stars reside in higher order gravity coupled with higher dimensions. Equivalently, higher mass stars are predicted by classical Einstein gravitation theory. The plots for compactness are depicted in Fig. \ref{com} which signify that all the solutions satisfy the Buchdahl limit $\frac{m}{r} < \frac{4}{9}$. Moreover, it is evident that for the $N = 2, d = 6 $ case (pure Gauss-Bonnet) the compactness ratio $\frac{m}{r}$ is the highest while for the Einstein case $N = 1, d = 4$ it is least.

\section{MATCHING OF INTERIOR AND EXTERIOR METRIC AT THE BOUNDARY}

The exterior metric is assumed to be a vacuum and is represented by
\begin{equation}
ds^2=-\left(1-{C_d \over r^{\frac{d-2N-1}{N}}}\right)dt^2+{dr^2 \over 1-C_d / r^{\frac{d-2N-1}{N}}}+r^2d\Omega^2_{d-2} \label{ext}
\end{equation}

where $C_d$ is given as
\begin{equation}
C_d = {8\pi G_d M \over d-2}~.~{\Gamma\left({d-1 \over 2}\right) \over \pi^{(d-1)/2}}
\end{equation}
with $G_d$, the gravitational constant in $d-$dimensions and $M$, the total mass of the stellar system as observed by an external observer.

On matching the metrics (\ref{int}) and (\ref{ext}) at the boundary $r=a$, we have
\begin{eqnarray}
e^{-\lambda_b} &=& 1-{C_d \over a^{\frac{d-2N-1}{N}}}=e^{\nu_b} \label{b1}\\
p_r(r=a) &=& 0. \label{b2}
\end{eqnarray}

Completing the matching of the interior and exterior geometries with these boundary conditions generates the values
\begin{eqnarray}
R &=& \sqrt{\frac{a^{d-1}-a^2 C_d}{ C_d}}\\
\nu_0 &=& \ln \left(1-{C_d \over a^{d-3}}\right)-\frac{(\alpha +1) a^2 (d-2 N-1)}{2 N R^2} \nonumber \\
&& \hspace{-0.5cm} -\alpha ~ ln \left(a^2+R^2\right) +\frac{\beta  a^{2 N+2} \left(\frac{a^2/R^2}{1+a^2/R^2}\right)^{-N-1}}{(d-2) N (N+1) R^2}\\
\frac{\beta }{\alpha } &=& \frac{(d-2)\big[a^2 (d-2 N-1)+(d-1) R^2\big]}{2(a^2+R^2)^{N+1}}  .
\end{eqnarray}
for these necessary constants and parameters. 
The remaining parameters  $d,~N,~a$ and $\alpha$ have been left free and a range of values will be examined to study the model properties.

\begin{figure}[htbp]
\includegraphics[scale=0.7]{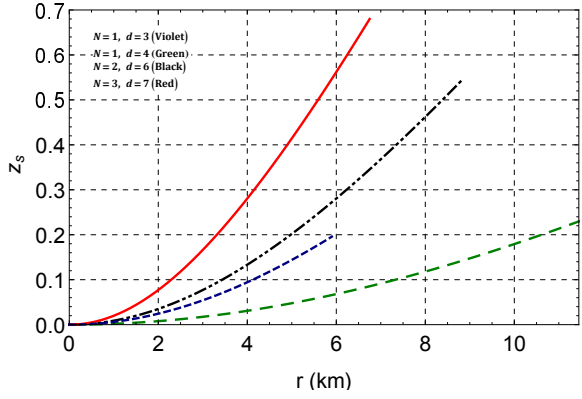}
\caption{\label{red}  Variation of gravitational red-shift with radial coordinates $r$ for the parameters given in Table \ref{tab}.   }
\end{figure}

Also the surface red-shift can be completely determined from
\begin{eqnarray}
z_s &=& e^{-\nu_s/2}-1=e^{\lambda_s/2}-1\\
&=& \sqrt{1+{a^2 \over R^2}}-1.
\end{eqnarray}
We observe that the surface redshift due to higher order effects is larger than their classical Einstein counterparts Fig. \ref{red}.  

\begin{figure}[htbp]
\includegraphics[scale=0.7]{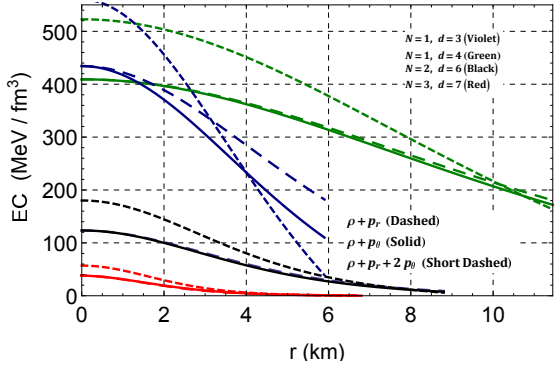}
\caption{\label{ener} Variation of energy conditions (EC) with radial coordinate $r$ for the parameters given in Table \ref{tab}.   }
\end{figure}

\section{ENERGY CONDITIONS}
The solutions of the field equations in Lovelock gravity can represent a physically viable stellar distribution only if it satisfies certain energy conditions. Here we are going to verify graphically that our solution hold the null energy condition (NEC), dominant energy condition (DEC), strong energy condition (SEC) and weak energy condition (WEC) conditions at all the interior points i.e.
\begin{eqnarray}
\mbox{NEC}&:&~~\rho(r) \ge 0\\
\mbox{WEC}&:&~~\rho(r)+p_r(r) \ge 0~~,~~\rho(r)+p_\theta(r) \ge 0\\
\mbox{SEC}&:&~~ \rho(r)+p_r(r)+2p_\theta(r) \ge 0\\
\mbox{DEC}&:&~~\rho(r) \ge |p_r(r)|,~|p_\theta(r)|.
\end{eqnarray}
which is verified in Fig. \ref{ener}. 

\section{EQUILIBRIUM AND STABILITY ANALYSIS}

\subsection{Equilibrium analysis}

\begin{figure}[htbp]
\includegraphics[scale=0.7]{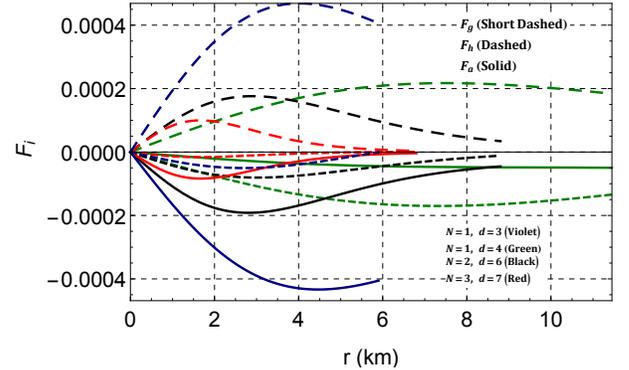}
\caption{\label{to1}  Variation of forces due to gravity ($F_g$), hydrostatic ($F_h$) and anisotropy ($F_a$) acting through Tolman-Oppenheimer-Volkoff equation with radial coordinates $r$ for the parameters given in Table \ref{tab}.   }
\end{figure}

For a stellar configuration to be stable, all the forces acting on the system has to counter-balance each other (or at equilibrium). The balanced force equation is given by the generalized TOV-equation in higher dimensions and it can be obtained from the conservation of energy-momentum tensor $T_{AB}$ given in (\ref{tov}) i.e.
\begin{eqnarray}
-{1 \over 2} (p_r+\rho){d\nu \over dr}-{dp_r \over dr}+{d-2 \over r}(p_\theta-p_r)=0
\end{eqnarray}
which further can reduce to 
\begin{eqnarray}
F_g+F_h+F_a=0 
\end{eqnarray}
where $F_g,~F_h$ and $F_a$ are defined in (\ref{fg})-(\ref{fa}).
The counter balancing of these force are shown in Fig. \ref{to1}. We should point out that the contributions from anisotropy can lead to  repulsive ($p_\theta > p_r$) or attractive ($p_\theta < p_r$) forces. For this case $p_r>p_\theta$ signifying an attractive anisotropic force. Hydrostatic equilibrium is achieved via the TOV-equation by the balancing of the hydrostatic force and the combined force due to anisotropy and gravity.

\begin{figure}[htbp]
\includegraphics[scale=0.7]{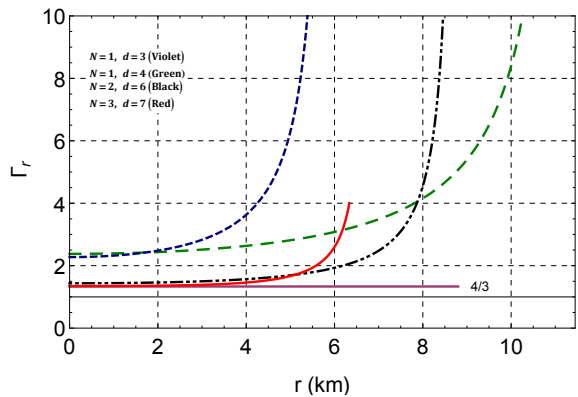}
\caption{\label{gamm} Variation of relativistic adiabatic index with radial coordinate $r$ for the parameters given in Table \ref{tab}.   }
\end{figure}

\subsection{Adiabatic index and stability}
The relativistic adiabatic index $\Gamma_r$ of the stellar configuration can be defined as \citep{cha93}
\begin{equation}
\Gamma_r = {p_r+\rho \over p_r} {dp_r \over d\rho}.
\end{equation}

Bondi \cite{bon64} has discussed that for a Newtonian fluid sphere obeying a polytropic equation of state (EoS), an adiabatic collapse will proceed if $\Gamma_r \le 4/3$ and the collapse will be catastrophic if less than $4/3$. Therefore, the stiffness of a Newtonian fluid sphere is not sufficient enough to hold the mass if the adiabatic index is less than $4/3$. 

However, for an anisotropic fluid sphere this condition is adapted for a relativistic fluid sphere due to its regenerative effect of pressure. For an anisotropic relativistic sphere to initiate adiabatic collapse \cite{cha93}, 

\begin{equation}
\Gamma<\frac{4}{3}+\left[\frac{4}{3}\frac{(p_{\theta 0}-p_{r0})}{|p_{r0}^\prime|r}+{8\pi \over 3} \frac{\rho_0p_{r0}}{|p_{r0}^\prime|}r\right]_{max},
\end{equation}
where, $p_{r0}$, $p_{\theta 0}$, and $\rho_0$ are the initial radial, tangential, and energy density in static equilibrium satisfying. The first and last terms inside the square brackets represent the anisotropic and relativistic corrections respectively and both the quantities are positive which increase the unstable range of $\Gamma$ \cite{cha93,cha92,her92}. Now we are convinced that for an anisotropic fluid, adiabatic collapse is still possible even if the adiabatic index is more than $4/3$ for positive values of anisotropy. However, this condition will change depending on the various types of anisotropy. Fig. \ref{gamm} shows that our model is stable at each interior point of the fluid configuration. Furthermore, higher order effects render the star more unstable than their classical Einstein counterparts since the central values of $\Gamma$ is much lower in higher $N$ and $d$.

\begin{figure}[htbp]
\includegraphics[scale=0.7]{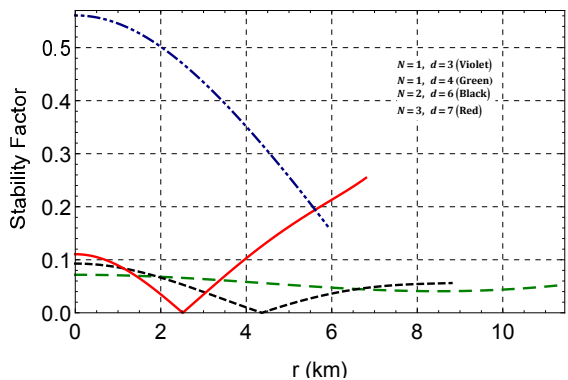}
\caption{\label{sta}  Variation of stability factor $|v_t^2-v_r^2|$ with radial coordinates $r$ for the parameters given in Table \ref{tab}.   }
\end{figure}

\begin{figure}[htbp]
\includegraphics[scale=0.7]{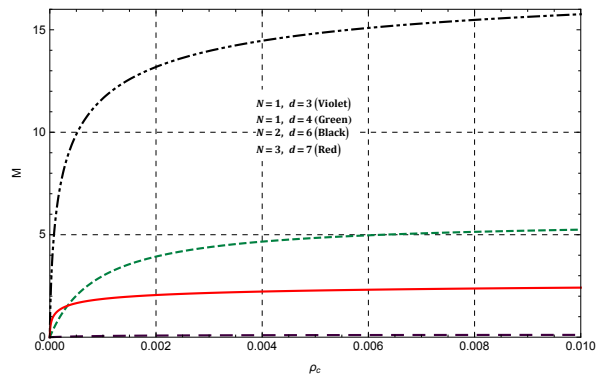}
\caption{\label{ma-ro}  Variation of mass with central density for the parameters given in Table \ref{tab}.}
\end{figure}

\subsection{Causality condition and stability}
For any physical fluids, satisfaction of the causality condition is required i.e. $0 \le (v_r^2,~v_\theta^2) \le 1$. Abreu et al. \cite{abr07} have shown that since both the components of sound speed are less than 1, their difference has to satisfy the inequality $0 \le |v_\theta^2-v_r^2| \le 1$. The parameter $|v_\theta^2-v_r^2|$ is defined as {\it stability factor} and it has to be less than 1 for a potentially stable configuration \cite{abr07}. We observe an interesting phenomenon in Fig \ref{sta} in which the stability factor decreases to zero for some finite radius and then increases towards the boundary. The vanishing of the stability factor occurs at a much smaller radius for higher order theories of gravity and the subsequent increase towards the surface is much sharper.

\subsection{Static stability criterion}
The concepts of radial perturbation was put forwarded by S. Chandrasekhar \cite{chand} where the metric functions and the physical parameters i.e. density and pressure were perturbed as 
\begin{eqnarray}
&& \lambda \rightarrow \lambda_0+\delta \lambda~,  ~ \nu \rightarrow \nu_0+\delta \nu~, \nonumber\\
&& \rho \rightarrow \rho_0+\delta \rho~,~   p\rightarrow p_0+\delta p. \nonumber
\end{eqnarray}
Then all the perturbations are taken as oscillatory function $\delta \lambda \cdot \exp[i\sigma t]$ etc. with $\sigma$ as characteristics frequency. Now, the amplitude of the perturbations  $\delta \lambda,~\delta \nu,~\delta \rho,~\delta p$ are determined from the perturbed field equations. Further, the values of the characteristic frequencies are determined from the conservation equation. If $\sigma^2<0$  or $\sigma \equiv \pm i\sigma$ one has collapsing/expanding (unstable) system and  $\sigma^2>0$  or $\sigma \equiv \pm \sigma$  leads to an oscillatory i.e. non-collapsing/expanding (stable) system.

This method has been simplified as the static stability criterion by Harrison et al. \cite{har65} and Zeldovich-Novikov \cite{zel71}. This suggests  that stability is always maintained as long as the mass of the configurations increase with central density or mathematically $dM/d\rho_c > 0$ for any stable configuration. The point where $dM/d\rho_c = 0$ is called the turning point. For this solution the mass in terms of central density can be written as
\begin{eqnarray}
M(\rho_c) &=& \frac{2 \pi ^{(d-1)/2}}{\Gamma \left(\frac{d-1}{2}\right)} \left[\frac{d-2}{16 \pi } a^{d-2 N-1} \left(\frac{a^2}{a^2+\tilde{\rho}_c^2}\right)^N\right] \\
{dM \over d\rho_c} &=& \frac{(d-2) \pi ^{\frac{d-3}{2}} a^{d-2 N-1} \left(\frac{(d-2) (d-1)}{\rho _c}\right)^{1/N}}{8 \rho _c \Gamma \left(\frac{d-1}{2}\right) \left(a^2 (16 \pi )^{1/N}+\left(\frac{(d-2) (d-1)}{\rho _c}\right)^{1/N}\right)} \nonumber \\
&& \left[\frac{a^2}{a^2+(16 \pi )^{-1/N} \left(\frac{(d-2) (d-1)}{\rho _c}\right){}^{1/n}}\right]^N > 0
\end{eqnarray}
where $\tilde{\rho}_c^2 = [(d-1)(d-2)/(2\rho_c)]^{1/N}$. The variation of mass with central density is shown in Fig. \ref{ma-ro}. Here the range of stable density during radial perturbation is highest for second order 6D-gravity (i.e. $N=2$) than the rest implying that compact stars are much stable under radial perturbations.

\begin{table*}
\caption{\label{tab} Values of all the parameters for well-behaved solutions.}
\begin{ruledtabular}
\begin{tabular}{llllllllllp{0.55in}}
$N$ & $d$ & $R$ & $\beta$ & $\alpha$ & $\nu_0$ & $a$ & $M_{max}$ & $z_s$ & $u={M_{max} \over a}$ & $S(r=a)$\\ 
&& $(km)$ & $(km^{-2})$ &&& $(km)$ & $(M_\odot)$ &&& $MeV/fm^2$\\
\hline
1 & 3 & 9 & $19.8 \times 10^{-4}$ & 0.33 & 0.01 & 3.88 & 0.98 & 0.19 & 0.373 & 618.08\\
1 & 4 & 16 & $19.8 \times 10^{-4}$ & 0.33 & 0.01 & 10.71 & 2.67 & 0.23 & 0.368 & 607.36\\
2 & 6 & 7.5 & $1 \times 10^{-4}$ & 0.33 & 0.01 & 2.26 & 0.499 & 0.54 & 0.326 & 87.166\\
3 & 7 & 5 & $5 \times 10^{-6}$ & 0.33 & 0.001 & 4.69 & 0.869 & 0.67 & 0.274 & 7.151\\
\end{tabular}
\end{ruledtabular}
\end{table*}

\begin{figure}[htbp]
\includegraphics[scale=0.7]{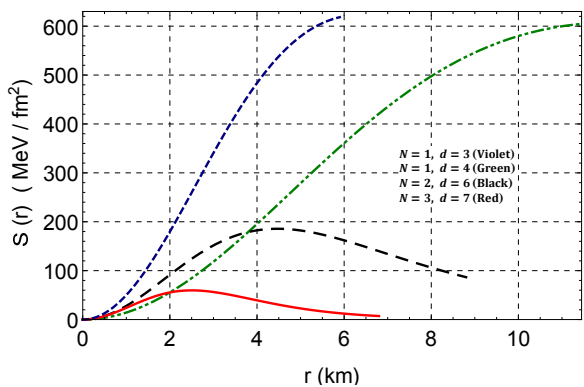}
\caption{\label{surfa} Variation of surface tension with surface radius for the parameters given in Table \ref{tab} along with $n=0.725 ~fm^{-3}$ or equivalently $r_n=0.76~fm$.  }
\end{figure}

\section{SURFACE TENSION OF COMPACT STARS AND ITS BEHAVIOUR IN HIGHER DIMENSIONS}

A recent investigation by Bagichi et al \cite{bag05} on the surface tension of compact stars reveal that the surface tension of ``strange stars" (composed of $u,~d,~s$-quarks) is higher than neutron stars. The surface tension of a non-rotating spherically symmetric self-gravitating strange matter can be calculated from the excess pressure on the surface given as
\begin{equation}
|\Delta p_r|_{r=a}={2S \over a} \label{surf}
\end{equation}
where $S$ is the surface tension and $a$ is the radius of the star. The excess pressure $\Delta p_r$ at the surface can be determined from
\begin{equation}
|\Delta p_r|_{r=a}= r_n \left|{dp_r \over dr}\right|_{r=a}.
\end{equation}

Here $r_n$ is the radius of quark particle and is given by $r_n=(1/\pi n)^{1/3}$, where $n$ is the baryon number density. By assuming the baryon number density $n=0.725~MeV/fm^2$, we have obtained the behavior of surface tension in higher dimensions. It is clearly seen from Fig. \ref{surfa} that the surface density decreases significantly with increase in higher order and dimensions. For $N=1, ~d=3,4$ it seems to increase the surface tension  with increasing surface radius of the compact star. However, in the case of higher order gravity in higher dimensions i.e. for $N=2,~d=6$ and $N=3,~d=7$ the surface tension is maximum for a particular value of surface radius at about 4.45 km and 2.48 km respectively. We observe that the surface tension seems to be directly proportional to the measure of anisotropy $\Delta(r)$. For $N=1, ~d=3,4$ the anisotropy is maximum at the surface and hence surface tension as well. Similarly, for $N=2,~d=6$ and $N=3,~d=7$, whenever the anisotropy increases the surface tension also increases and vice-versa. According to Alcock and Olinto \cite{alc} the existence of strange stars requires a large value of $S$ which can be obtained by general relativistic corrections in modeling such compact stars \cite{bag05}. However, Sharma and Maharaj \cite{sha07} have shown that a wide range in the values of $S$ can be obtained by introducing pressure anisotropy. In this article, we have provided a mechanism which accounts for wide ranges of $S$ by incorporating higher order gravity in higher dimensions.

\begin{figure}[htbp]
\includegraphics[scale=0.7]{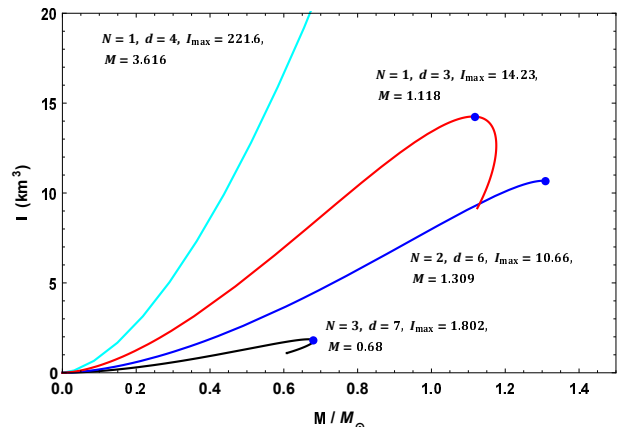}
\caption{\label{im} Variation of $I$ with $M/M_\odot$ for different dimension and order of gravity.}
\end{figure}

\section{Slow rotation in different dimensions and higher order gravity}

For a rotating compact star, the stiffness of the EoS is more sensitive to the maximum value of the moment of inertia $(I_{max})$ than $M_{max} ~or ~R_{max}$, Haensel et al. \cite{hae}. Therefore, it is necessary to calculate the moment of inertia. Assuming a rigidly rotating star at angular frequency $\Omega$, its moment of inertia can be determined as \cite{latt}
\begin{eqnarray}
I = {8\pi \over 3} \int_0^R r^4 (\rho+p_r) e^{(\lambda-\nu)/2} ~{\bar{\omega} \over \Omega}~dr
\end{eqnarray}

where, the local spin frequency $\bar{\omega}$ satisfies Hartle's equation \cite{hart}
\begin{eqnarray}
{d \over dr} \left(r^4 j ~{d\bar{\omega} \over dr} \right) =-4r^3\bar{\omega}~ {dj \over dr} .
\end{eqnarray}
with $j=e^{-(\lambda+\nu)/2}$ which has boundary value $j(R)=1$. The approximate moment of inertia $I$ up to the maximum mass $M_{max}$ was given by Bejger and Haensel \cite{bejg} as
\begin{equation}
I = {2 \over 5} \Big(1+x\Big) {MR^2},
\end{equation}
where parameter $x = (M/R)\cdot km/M_\odot$. For the solution we have plotted $M/M_\odot$ vs $I$ in Fig. \ref{im}. Here we have found that the maximum moment of inertia $I_{max}$ occurs for $N=1$ and $d=4$ (the Einstein's gravity). For Einstein's gravity in 3-D, the $I_{max}$ decreases as $M_{max}$ decreases. For EGB gravity ($N=2,~d=6$) the values of $I_{max}$ and $M_{max}$ are even lower than the values in Einstein's gravity. These values of $I_{max}$ and $M_{max}$ further decrease in Lovelock gravity ($N=3,~d=7$). These observations imply that the compact stars can attain maximum stiffness in EoSs in Einstein's gravity with the rotation and mass being maximum. In fact the mass corresponding to $I_{max}$ is lower by $\sim 3$\% from the $M_{max}$. This happens to the EoS without any strong high-density softening due to hyperonization or phase transition to an exotic state \cite{bej}.

\section{RESULTS AND DISCUSSIONS}

In this paper we have constructed exact models of compact objects within the framework of pure Lovelock gravity. We assumed that the radial and tangential stresses at each interior point of the stellar fluid are unequal. Furthermore, the interior matter distribution obeys a linear equation of state which relates the radial pressure to the matter density. In order to close the system of equations we assume a form for the one of the gravitational potentials based on the Finch and Skea ansatz. The equation of state readily gives the other metric potential thus completing the gravitational behaviour of the model. A thorough investigation of the physical viability of the model based on thermodynamical properties and stability analysis indicate that our model is an excellent star-like candidate residing in pure Lovelock gravity. In a recent paper by Dadhich et al. \cite{dad} exact solutions in pure Lovelock gravity for isotropic perfect fluid distributions were presented. An interesting observation arising from this work is that there cannot exist bounded configurations describing star-like objects for odd dimensions ($d=2N+1$)  \cite{hans-gab}.   The same is not true for even dimensions ($d=2N+2$) \cite{hans-gab2}. This arises from the fact that the radial pressure does not vanish for some finite $r$ which defines the boundary of the object. In the present work we have found bounded configurations for odd dimensions but by relaxing to the anisotropic case and by carefully choosing the equation of state of the stellar fluid. We must point out that the boundedness of the solutions does not arise from the fact that the stresses within the fluid configuration are unequal (ie., $p_r \neq p_\theta)$ but from the nature of the EoS which assumes the form $p_r = \alpha\rho - \beta$ where $\beta = \alpha\rho_s$. Here $\rho_s$ is the energy density at the stellar surface. One concludes that this choice of the EoS guarantees the vanishing of the radial pressure for some finite value $r > 0$. Thus within this construction, it is possible to have bounded configurations for both odd and even dimensions. 
It is already well known in the literature that compact star solutions exist in Einstein-Gauss-Bonnet gravity which is the case of $N=2$ and $d=5$.

\section{Conclusions}

An exact model of an anisotropic star in the context of pure Lovelock gravity was obtained by prescribing a linear equation of state and a Finch-Skea potential. The general solution was found for all spacetime dimensions $d$ and order of Lovelock polynomial $N$.  Consequently it was possible to compare the effect of $d$ and $N$ on the models. It was found that the Einstein star was the least compact and that the mass-radius ratio decreased with increasing $d$ and $N$. Further Einstein stars generated the highest density and pressure values. The results of this investigation show decidedly that higher curvature effects have a marked influence on stellar structure when compared to stars in the standard theory general relativity.  

\section*{Acknowledgement}
We thank Professor L. Herrera for providing us with critical insights on anisotropic fluids. We are also thankful to the esteemed reviewers for their valuable suggestions that greatly improved the manuscript.

\end{document}